\documentclass[final,3p,twocolumn,sort&compress]{elsarticle}
\usepackage{balance}
\usepackage{chngcntr}
\usepackage{cuted} 
\usepackage{lipsum} 
\usepackage{natbib}
\usepackage{ulem}
\usepackage{amsmath}
\usepackage{appendix}
\usepackage{url}
\usepackage[colorlinks=true,linkcolor=blue,filecolor=blue, urlcolor=blue,citecolor=blue]{hyperref}
\usepackage{float}
\usepackage[T1]{fontenc}
\usepackage{graphicx}
\usepackage{mathrsfs}
\usepackage{dcolumn}
\usepackage{bm}
\usepackage{cases}
\usepackage{multirow,booktabs}
\usepackage{makecell}
\usepackage{amssymb}
\usepackage{eucal}
\usepackage{subfigure}
\usepackage{natbib}
\usepackage{amsthm,amsmath,amssymb}
\usepackage{mathrsfs}
\usepackage{subeqnarray}
\usepackage{color}
\usepackage{ulem}
\usepackage{soulutf8}
\usepackage{appendix}
\usepackage{listings}
\usepackage{braket}
\usepackage{multirow,booktabs}
\usepackage{chngcntr}

\begin{document}

\begin{frontmatter}



\title{Universal laws for nuclear contacts}

\author[1]{Tongqi Liang}

\author[2]{Dong Bai}

\author[1,3]{Zhongzhou Ren\corref{cor1}}
\ead{zren@tongji.edu.cn}



\cortext[cor1]{Corresponding author}
\address[1]{School of Physics Science and Engineering, Tongji University, Shanghai 200092, China}
\address[2]{College of Mechanics and Engineering Science, Hohai University, Nanjing 211100, China}
\address[3]{Key Laboratory of Advanced Micro-Structure Materials, Ministry of Education, Shanghai 200092, China}

%

\begin{abstract}

The nuclear contact characterizes the nucleon-nucleon pairs in close proximity and serves as an important tool for studying the short-range correlations (SRCs) within atomic nuclei. While they have been extracted for selected nuclei, the investigation of their behavior across the nuclear chart remains limited.
Very recently, Yankovich, Pazy, and Barnea have proposed a set of universal laws (YPB laws) to describe the correlation between nuclear contacts and nuclear radii and tested their laws for a small number of nuclei by using the Woods-Saxon mean-field model~[R.\ Yankovich, E.\ Pazy, and N.\ Barnea, arXiv:2407.15068 (2021)]. In this Letter, we extend their study to a majority part of the nuclear chart within the framework of the Skyrme Hartree-Fock-Bogolyubov model, which incorporates several essential beyond-mean-field features and offers a more accurate description of the bulk properties of atomic nuclei. Our results suggest that the YPB laws hold as a good approximation for different nuclear mass regions, with minor deviations attributed to, e.g., isospin-breaking effects. Our work lays a firm foundation for future applications of the YPB laws in finite nuclei and provides new evidence for the long-range nature of the relative abundance of short-range pairs.

\end{abstract}

\end{frontmatter}
	
\section{ Introduction}
A complete theoretical description of the nuclear many-body wave function is a complex task in nuclear physics, particularly when focusing on the short-distance and high-momentum components~\cite{Ciofi2015,Liang2022,Niu2022,Liang20241}.
These components arise from the strongly interacting nucleon pairs driven by the short-range correlations (SRCs), which have implications for neutron star properties~\cite{Hen2015,cai2016,Lu2022,Hong2023}, European Muon Collaboration (EMC) effects~\cite{Weinstein2011,Hen2012,Hen2017,chen2017}, and neutrinoless double beta decay matrix elements~\cite{Simkovic2009,Cirigliano2018,Jokiniemi2021,Weiss2022}.  

The generalized contact formalism (GCF), inspired by the approach to SRCs in dilute Fermi systems, has garnered significant success in the study of nuclear SRCs~\cite{TAN20082952,TAN20082971,TAN20082987,Hen20152, Weiss2015}.
In GCF, the nuclear contacts, characterizing the probability of finding nucleon pairs in close proximity, are shown to be related to many nuclear quantities, such as high-momentum tails, two-body densities in coordinate space, and the Levinger constant in the photo-absorption cross section~\cite{Cruz-Torres2020,Weiss20152,Weiss2018}. These relationships have enabled the extraction of nuclear contacts for chosen stable nuclei from quasi-elastic electron scattering experiments and for light nuclei from ab-initio many-body calculations~\cite{Weiss2018}. 
Recently, the BM@N Collaboration measured the quasi-free scattering of $^{12}$C ions from hydrogen, showcasing the potential to explore nuclear contacts for unstable nuclei~\cite{Patsyuk2021}. 
However, due to experimental limitations and high computational costs, research on nuclear contacts has largely been confined to stable and light nuclei.

The contact coefficient ratio between two different nuclei has been shown to remain unchanged for various interaction models, indicating that the relative abundance of short-range pairs is a long-range quantity, insensitive to the specific details of nucleon-nucleon short-distance interaction~\cite{Cruz-Torres2020}. Our previous studies revealed a model-independent correlation between the proton-proton contact and the root-mean-square (RMS) radius of proton density distributions in Sn isotopes~\cite{Liang2024}. In Ref.~\cite{Yankovich2024}, based on the Woods-Saxon mean-field description, Yankovich, Pazy, and Barnea proposed that nuclear contacts and RMS radius of nucleon density distributions exhibit universal correlations across all nuclei and they are linked by the so-called generalized Levinger constants, which we refer to as the YPB laws.
The YPB laws establish a connection between SRCs and RMS radius of nucleon density distributions, a long-range nuclear property, offering new insights into the formation of SRC pairs.
Preliminary verification of YPB laws has been achieved through the application of the Woods-Saxon mean-field model for a limited number of nuclei. 
They have proposed that, at leading order, the generalized Levinger constants are independent of the specific nucleus, and generalized spin-zero Levinger constants are equal due to isospin symmetry, approximately.

In this letter, we combine the Skyrme Hartree-Fock-Bogolyubov (HFB) model and GCF to validate the YPB laws and constrain the generalized Levinger constant.
The calculations are extended to all nuclei across the chart of nuclides and global generalized Levinger constants are evaluated, which offers a practical method to evaluate nuclear contacts from nuclear radii.
Within the framework of the Skyrme HFB model and GCF, we assess the isospin-breaking effects introduced by the Coulomb force, leading to modifications of the YPB laws and providing a more robust foundation for their future application.
The neutron skin—the difference between neutron and proton matter radii—provides a crucial constraint on the symmetry energy in the equation of state (EOS)~\cite{Centelles2009,Adhikari2021,Adhikari2022,Reinhard2021,Reinhard2022,Liang2023}. Additionally, short-range correlations (SRCs) significantly influence neutron star properties. The findings in this study suggest a potential new perspective on the interconnections between the neutron skin, SRCs, and the EOS, highlighting the broader implications of the YPB laws and their role in linking short-range nuclear interactions with long-range nuclear properties.




\section{ Method}\label{II}

The GCF is based on the scale separation between the strong short-range interaction in SRC pairs and their interactions with the residual $A-2$ system in nuclei. Under this assumption, the many-body wave function of nuclei can be factorized into a product of an asymptotic two-body wave function describing the underlying short-range behavior and a function describing the surrounding nucleons.
The contributions of the residual $A-2$ nucleons to the SRC pairs collapse into coefficients called ``nuclear contacts''. As the universal short-range behavior is described by the asymptotic two-body wave function, at short separation distance ($r=|\boldsymbol{r}|<r_0\approx 0.9$ fm), the two-body pair-density $\rho^A_{\alpha,N_1\!N_2}(\boldsymbol{r})$, which represents the probability for finding a nucleon-nucleon pair, can be modeled by
\begin{equation}\label{contact1}
    \rho^A_{\alpha,N_1\!N_2}(\boldsymbol{r})=C^A_{\alpha,N_1\!N_2}\times|\varphi^{\alpha}_{N_1\!N_2}(\boldsymbol{r})|^2,
\end{equation}
where $C^A_{\alpha,N_1\!N_2}$ is the nuclear contact and $\varphi^{\alpha}_{N_1\!N_2}(\boldsymbol{r})$ is the asymptotic two-body wave function. $N_1\!N_2$ can be proton-proton ($pp$), neutron-neutron ($nn$), and neutron-proton ($np$) pairs. $\alpha$ marks the channel of two correlated nucleons. In GCF, only the spin-one $np$ channel and spin-zero $nn$, $pp$, and $np$ channels are considered, as higher partial waves are suppressed~\cite{Weiss2019}.

In Ref.~\cite{Yankovich2024}, Yankovich, Pazy, and Barnea proposed that spin-zero nuclear contacts can be evaluated with the form:
\begin{equation}
    C^A_{0,N_1 N_2}\sim\rho_{N_1} \rho_{N_2} \Omega_{N_1\!N_2},
\end{equation}\label{Levinger1}
where $\rho_{N_1}$ and $\rho_{N_2}$ are nucleon densities and $\Omega_{N_1\!N_2}$ represents the relevant nuclear volume.
Considering uniform distribution densities, nucleon volume is estimated as $\Omega_{p(n)} \sim R_{p(n)}^3$ with $R_{p(n)}$ the RMS radius of proton (neutron) density distributions, leading to the proton (neutron) density $\rho_{p(n)}\sim {Z(N)}/R_{p(n)}^3$, where $Z(N)$ is the proton (neutron) number. Under these approximations, the relationship in Eq.~(\ref{Levinger1}) can be rewritten as
\begin{equation}\label{Levinger2}
    \begin{aligned}
        C^A_{0,pp}&=L^0_{pp}\frac{Z^2}{R_p^3}, \\
        C^A_{0,nn}&=L^0_{nn}\frac{N^2}{R_n^3}.
    \end{aligned}
\end{equation}
$L^0_{pp}$ and $L^0_{nn}$ are the generalized Levinger constants. 
In this work, the above correlations between nuclear contacts and nuclear radii are referred to as YPB laws.
As for $np$ pairs, there are two channels: the spin-zero channel and spin-one channel (the deuteron channel) induced by the tensor force. The $np$ two-body density distributions are expressed as: $\rho^A_{np}(r)=C^A_{0,np}|\varphi^0_{np}(r)|^2+C^A_{1,np}|\varphi^1_{np}(r)|^2$. The relative amplitude of $C^A_{0,np}$ and $C^A_{1,np}$ is difficult to determine and universal laws for $np$ pairs are not available in this work.

To verify the relationships mentioned in Eq.~(\ref{Levinger2}), it is necessary to simultaneously evaluate both nuclear contacts and RMS radii. One intuitive method to address spin-zero nuclear contacts is to compare the two-body wave function $\varphi^0_{N_1\!N_2}(\boldsymbol{r})$ to the two-body pair-density $\rho^A_{0,N_1\!N_2}(\boldsymbol{r})$. However, ab-initio calculations of two-body pair-densities with realistic interactions are typically restricted to the light and medium nuclear regions due to their significant computational demands~\cite{Hu2022,Lonardoni2017}. Following the studies in Refs.~\cite{Weiss2019,Liang2024}, we explore an approach capable of evaluating two-body pair-densities and then nuclear contacts across various isotopic and isotonic chains.

At large separation distances, correlations between two nucleons are expected to be unimportant. Therefore, one can construct $\rho^A_{0,N_1\!N_2}(\boldsymbol{r})$ by integrating the one-body point-nucleon density $\rho_N(\textit{\textbf{r}})$
\begin{equation}\label{onebody}
\begin{aligned}
    \rho^A_{0,N_1\!N_2} & (\boldsymbol{r})  \propto  \rho_{N_1\!N_2}^{U C}(\boldsymbol{r})\! \\
    & \equiv\! \int d \boldsymbol{R} \rho_{N_1}(\boldsymbol{R}+\boldsymbol{r} / 2) \rho_{N_2}(\boldsymbol{R}-\boldsymbol{r} / 2).
\end{aligned}
\end{equation}
For $pp$ and $nn$ pairs, the fermionic nature has to be considered, as two protons or two neutrons cannot occupy the same quantum state. Accounting for this exclusion, the two-body pair-density $\rho^A_{0,N_1\!N_2}(\boldsymbol{r})$ at large separation distances has the following form~\cite{Cruz2018}
\begin{equation}\label{twobody}
\begin{aligned}
        \rho^A_{0,N_1\!N_2}&(r)\! \underset{r \rightarrow \infty}{\longrightarrow}  \rho_{N_1\!N_2}^{F}(r) \\
        &\equiv \! \mathcal{N} \rho_{N_1\!N_2}^{U C}(r)\left[1-\frac{1}{2}\left(\frac{3 j_{1}\left(k_{F}^{} r\right)}{k_{F}^{} r}\right)^{2}\right],
\end{aligned}
\end{equation}
where $\mathcal{N}$ is the normalization factor, $j_1$ is the spherical Bessel function, and $k_F$ is the Fermi momentum of nucleons. The above relation is deduced from the expansion of the density matrix~\cite{Negele1972}, which offers a reliable approximation for nuclear matter and heavy nuclei. 
$\rho^F_{N_1\!N_2}(r)$ is considered to be a reasonable approximation for $\rho^A_{0,N_1\!N_2}(r)$ at large separation distances ($r>r_0$) because SRCs can be ignored in this region. The condition has been confirmed in Ref.~\cite{Weiss2019} by comparing $\rho^F_{pp}(r)$ with the two-body $pp$ pair-density of the VMC calculations.

At large separation distances where SRCs are not significant, Eqs. (\ref{onebody}) and (\ref{twobody}) provide an asymptotic expression for $\rho^A_{0,N_1\!N_2}(\boldsymbol{r})$. It enables the construction of $\rho^A_{0,N_1\!N_2}(\boldsymbol{r})$ directly from the one-body point-nucleon density.
In Eq.~(\ref{contact1}), we have demonstrated that $\rho^A_{0,N_1\!N_2}(\boldsymbol{r})$ at small separation distances can be well reproduced by nuclear contacts. 
Combining the short- and long-range behaviors of $\rho^A_{0,N_1\!N_2}(\boldsymbol{r})$, it is expected that both the nuclear contact and $\rho_{N_1\!N_2}^{F}(r)$ expressions can well describe the full two-body pair-density $\rho^A_{0,N_1\!N_2}(\boldsymbol{r})$, and the conclusion has been verified for $pp$ pairs in Ref.~\cite{Weiss2019}.
This indicates the extraction of spin-zero nuclear contact with the relation
\begin{equation}\label{CNN}
C^A_{0,N_1\!N_2}=\frac{\rho_{N_1\!N_2}^{F}\left(r_{0}\right)}{\left|\varphi_{N_1\!N_2}^{0}\left(r_{0}\right)\right|^{2}}.
\end{equation}
Based on the above formulas, the spin-zero nuclear contact can be obtained with only the one-body point-nucleon density $\rho_N(\boldsymbol{r})$.
 

We generate the point-proton and point-neutron density distributions using the Skyrme Hartree-Fock-Bogolyubov (HFB) method~\cite{Stoitsov2005,Sarriguren2007,Wang2020,Marevic2021}. Based on the Skyrme force, the variation of the HFB energy results in the Skyrme HFB equations.
The HFB method incorporates the pairing correlations as a necessary component for the description of open-shell nuclei.
Solving the HFB equations yields the quasiparticle states of constituent
neutrons and protons. The axially symmetric density distributions in the coordinate space are given by the sum of all quasiparticle states
\begin{equation}
        \rho_N(r, z)=\sum_k\left(\left|V_k^{+}(r, z)\right|^2+\left|V_k^{-}(r, z)\right|^2\right),
\end{equation}
where $V_k^{+(-)}(r, z)$ is the wave function corresponding to the positive quasiparticle energy $E_k$ and $(r,z)$ are the standard cylindrical coordinates. Then the RMS radii of neutron and proton density distributions are calculated with the expression
\begin{equation}
    R_{n,p}=\left(\frac{\int {r}^2 \rho(\boldsymbol{r})\text{d}\boldsymbol{r}}{\int \rho(\boldsymbol{r})\text{d}\boldsymbol{r}}\right)^{1/2}.
\end{equation}



\section{ Numerical results}

\begin{figure*}[h] \centering
\subfigure{
\label{Pb:a}	\includegraphics[width=0.9\columnwidth]{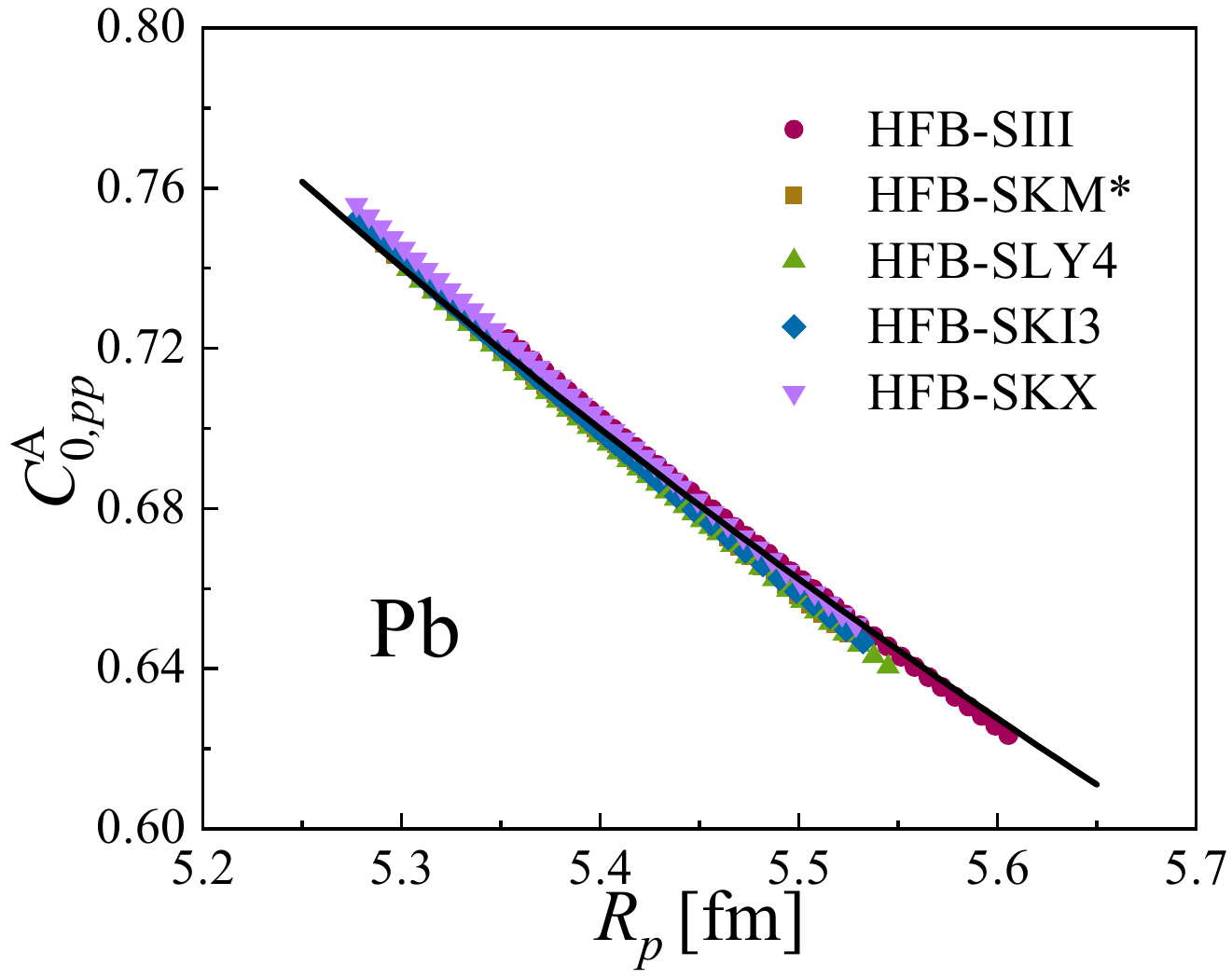}
		}
\subfigure {
\label{Pb:b}
\includegraphics[width=0.9\columnwidth]{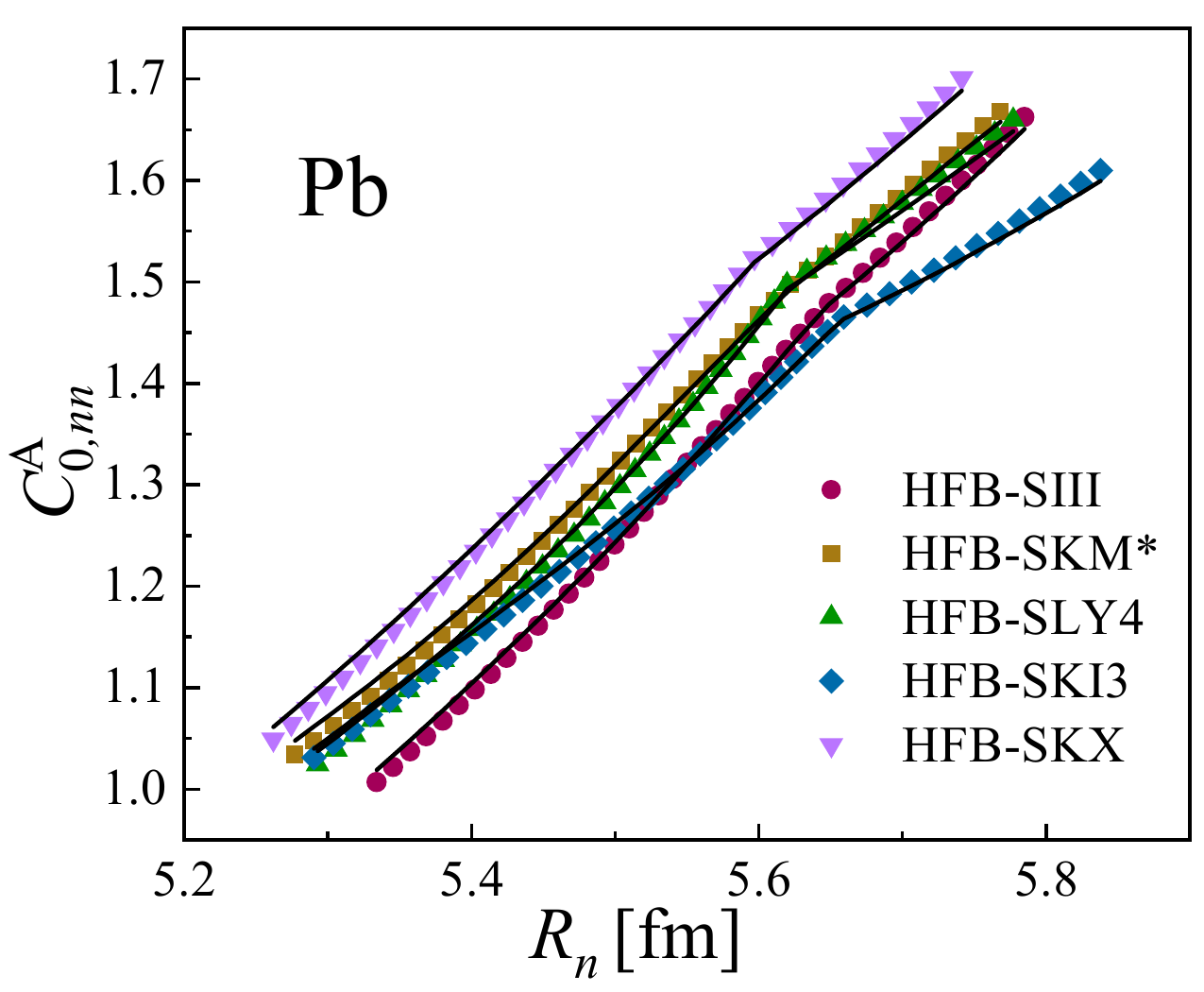}
		}
\caption{Nuclear contacts for Pb isotopes versus RMS radii of nucleon density distributions for (a) $pp$ pairs and (b) $nn$ pairs. The results are shown for nuclei with neutron numbers ranging from $N=96$ to $N=138$. The fitted lines follow the expressions $C^A_{0,pp}=L^0_{pp}\frac{82^2}{R_p^3}$ and $C^A_{0,nn}=L^0_{nn}\frac{N^2}{R_n^3}$. The corresponding point-nucleon density distributions are given by the Skyrme HFB method with SIII, SKM*, SLY4, SKI3, and SKX parameter sets.}
\label{Pb}
\end{figure*}



\begin{figure*}[h] \centering
\includegraphics[width=1.9\columnwidth]{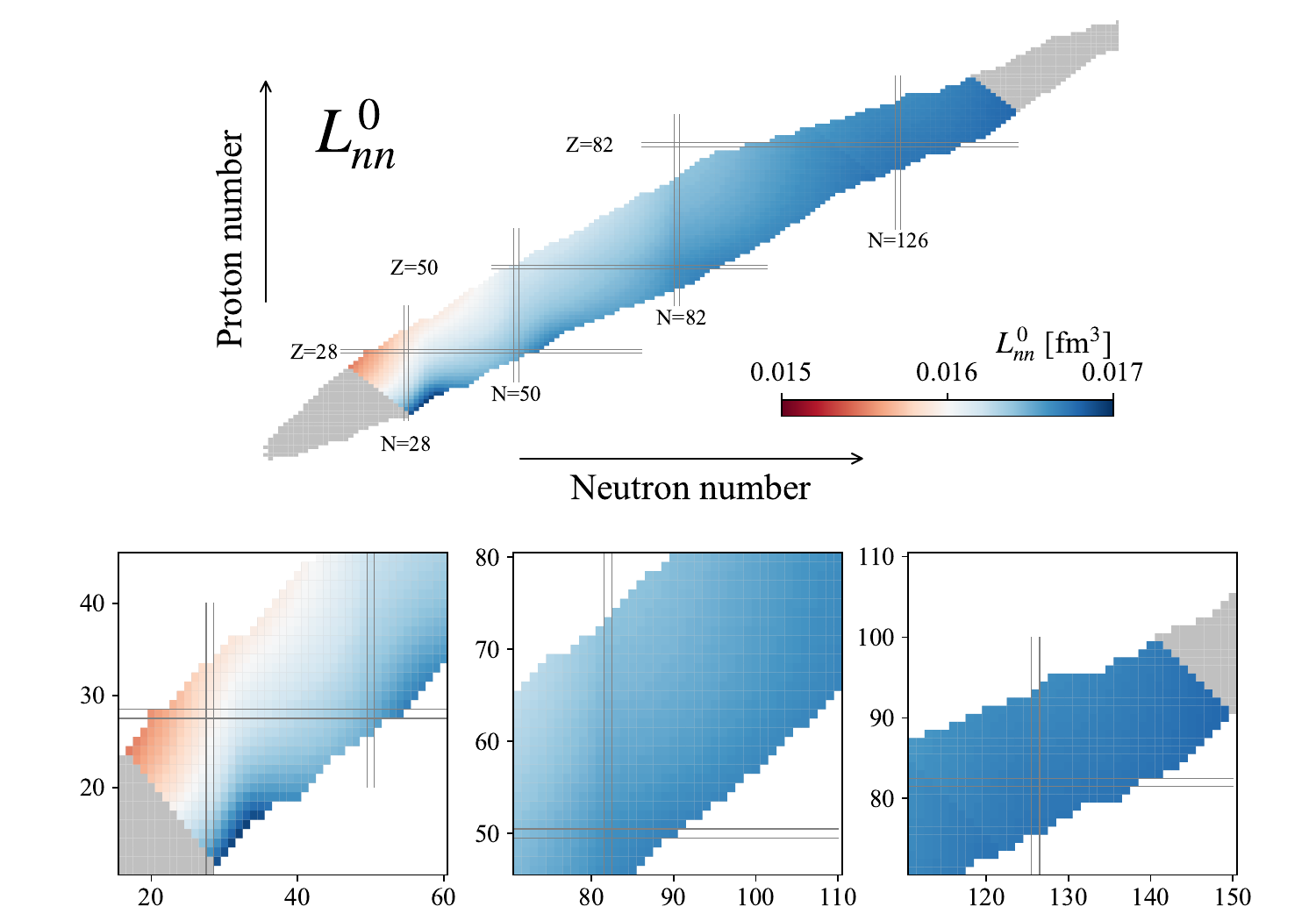}
\caption{Generalized Levinger constant for $nn$ pairs $L^0_{nn}$ across the chart of nuclides. The three images below display an enlarged view of the pattern. The corresponding $nn$ contacts and RMS radii of neutron density distributions are calculated with the HFB point-neutron density distributions using the SLY4 parameter set.}
\label{L_nn}
\end{figure*}

With GCF and Skyrme HFB method shown in Sec.~\ref{II}, we try to verify the YPB laws across the nuclear chart, i.e. the universal relationships between nuclear contacts and nuclear radii. 
As shown in Eq.~(\ref{CNN}), $\varphi^{0}_{N_1\!N_2}(r_0)$ influences the absolute value of nuclear contacts, but it does not alter the universal laws of nuclear contacts because its impact is consistent across all nuclei.
In this work, the asymptotic two-body wave function $\varphi^{0}_{N_1\!N_2}(r)$ is obtained by solving the two-body Schrödinger equation at zero energy using the AV18 potential.  $\varphi^{0}_{N_1\!N_2}(r)$ is normalized such that the integral of its Fourier transform, $\tilde{\varphi}^{0}_{N_1\!N_2}(k)$, over momenta above the Fermi momentum $k_F$ is equal to one~\cite{Weiss2019}. In this way, the nuclear contacts demonstrate the probability of finding a nucleon-nucleon pair with relative momenta above $k_F$. 

In GCF, the uncertainty of nuclear contacts arises from the chosen matching point $r_0$ and the realistic potential for calculating $\varphi^{0}_{N_1\!N_2}(r)$, and the use of the infinite nuclear-matter approximation. The uncertainty is expected to be of the order of 10$\%$-20$\%$. However, the relationship between nuclear contacts and RMS radii remains unaffected by the uncertainty. In the following analysis, uncertainties are not included. The detailed results regarding uncertainties of nuclear contacts can be found in Refs.~\cite{Weiss2019,Liang2024}

In Fig.~\ref{Pb}, we present the nuclear contacts of the spin-zero $s$-wave channel of 43 Pb isotopes ($N=96-138$) as a function of nuclear radii. 
The point-nucleon density distributions are calculated by the Skyrme HFB method with five parameter sets: SIII, SKM*, SLY4, SKI3, and SKX).
For $pp$ pairs shown in Fig.~\ref{Pb:a}, the $pp$ contacts $C^A_{0,pp}$ consistently decrease with the increase of RMS radii of proton density distributions $R_p$. This indicates that protons are more likely to form SRC $pp$ pairs in regions of higher number density of nucleons. 
Although the values of $R_p$ vary depending on the choice of parameter sets, as do the results of $C^A_{0,pp}$ shown in the supplemental materials, the correlations between $C^A_{0,pp}$ and $R_p$ remain model-independent.
The $pp$ contacts $C^A_{0,pp}$ follow a clear $ \frac{1}{R^3_p}$ dependence, as predicted by Eq.~(\ref{Levinger2}) as the proton number is fixed to $Z=82$.
Using Eq.~(\ref{Levinger2}), we fit all these five data sets of Pb isotopes, and from this, we determine the generalized Levinger constant for $pp$ pairs $L^0_{pp}=0.01639$ fm$^3$. The corresponding fitting results are shown in the figure, which show very consistent results with HFB calculations.

In Fig.~\ref{Pb:b}, the correlations between $nn$ contacts $C^A_{0,nn}$ and RMS radii of neutron density distributions $R_n$ are presented. As $R_n$ increases, the $C^A_{0,nn}$ results for these five parameter sets show a consistent increase, although with different patterns. 
The solid lines represent the fitting results based on the expression of Eq.~(\ref{Levinger2}): $C^A_{0,nn}=L^0_{nn}\frac{N^2}{R_n^3}$, which closely match the calculated HFB $C^A_{0,nn}$ values.
Notably, the generalized Levinger constant $L^0_{nn}$ remains nearly identical for various parameter sets, ranging from 0.01669 to 0.01678 fm$^3$. 

The correlations presented in Fig.~\ref{Pb} validate the YPB laws that the nuclear contact is a function of nucleon number and nuclear radius, linked by the so-called generalized Levinger constants.
The YPB laws are also observed for $N=82$ isotones, as detailed in the supplemental materials, and for Sn isotopes, as shown in Ref.~\cite{Liang2024}.  By leveraging the YPB laws, the generalized Levinger constant can serve as a valuable tool for investigating nuclear contacts. It bridges the nuclear contact, a short-range nuclear property, to the long-range, mean-field nuclear radii, providing new insights into nuclear structure.

We calculate the nuclear contacts and extract generalized Levinger constants for all nuclei with mass number $40<A<240$. Lighter and heavier nuclei are excluded due to potentially inaccurate descriptions of the Skyrme HFB method. Shown in Fig.~\ref{L_nn} are the results of generalized Levinger constants for $nn$ pairs, $L^0_{nn}$, with the three images below offering an enlarged view of the pattern. It is noteworthy that the values of $L^0_{nn}$ range from 0.0155 to 0.0169 fm$^3$, remaining nearly constant across all nuclei. The consistency in $L^0_{nn}$ suggests a stable ratio between $nn$ contacts $C^A_{0,nn}$ and RMS radius of neutron density distribution. 
With the increasing mass number, the value of $L^0_{nn}$ slightly increases. 
Comparing the data for heavy nuclei, the variations of $L^0_{nn}$ among different nuclei are more pronounced for light nuclei. 
This phenomenon could be attributed to changes in the density distributions and the neutron-to-proton ratio, which fluctuate significantly in lighter nuclei.
Moreover, $L^0_{nn}$ values of neutron-rich nuclei are larger than those of proton-rich nuclei.
This observation indicates that compared to nuclei near the neutron drip line, the dependence of $C^A_{0,nn}$ on $\frac{N^2}{R_n^3}$ is weaker for nuclei near the proton drip line.
We also investigate the generalized Levinger constant for $pp$ pairs, with results detailed in the supplemental materials. The $L^0_{pp}$ results show a similar pattern as $L^0_{nn}$ but with slightly lower values ranging from 0.0151 to 0.0165 fm$^3$. Moreover, the $L^0_{nn}$ values also increase across the chart of nuclides. However, different from the $L^0_{nn}$ results, the $L^0_{pp}$ values for proton-rich nuclei are larger than those for neutron-rich nuclei. 

Due to isospin symmetry, it is expected that the generalized Levinger constants are equal for spin-zero $pp$ and $nn$ pairs for a symmetric $N=Z$ nucleus~\cite{Yankovich2024}. This implies that the nuclear contacts for both spin-zero $pp$ and $nn$ pairs should exhibit the same dependence on nuclear radii in these nuclei. The differences between $L^0_{pp}$ and $L^0_{nn}$ are shown in the supplemental materials. For proton-rich nuclei, $L^0_{pp}$ is larger than $L^0_{nn}$, while $L^0_{pp}$ has smaller values for neutron-rich nuclei. The same $L^0_{pp}$ and $L^0_{nn}$ results occur at the nuclear region where there are slightly more neutrons than protons.
This discrepancy may originate from the repulsive nature of the electromagnetic force between protons, which affects the point-nucleon densities and two-body wave functions, ultimately leading to differences in the generalized Levinger constants.

Based on the above analysis, the generalized Levinger constant for $NN$ pairs being composed of majority nucleons has larger values.
However, the generalized Levinger constant varies slightly. Therefore, it is reasonable to extract the global generalized Levinger constants for $nn$ pairs and $pp$ pairs, respectively.
We fit the data for all these nuclei and suggest the global generalized Levinger constants $L^0_{nn}=0.01659$ fm$^3$ and $L^0_{pp}=0.01626$ fm$^3$. 

The effects of tensor force cannot be accessed in YPB laws. With the increasing number of neutrons (protons), tensor force induces more $np$ pairs and leads to fewer $pp~(nn)$ pairs~\cite{Hen2014,Korover2021}. This aligns with the YPB laws that smaller nuclear radii, i.e., higher number density of nucleons, lead to larger spin-zero nuclear contacts. The competition between tensor force and the number density of nucleons requires further investigation.



\section{Conclusions}
\label{Concl}
Nuclear radii and nuclear contacts are governed by the long- and short-range part of the nuclear force, respectively. 
These two quantities are connected by the YPB laws, with the coefficient between them given by the generalized Levinger constant. 
For Pb isotopes and $N=82$ isotones, we find well-defined, model-independent relationships between nuclear contacts and nuclear radii for both $pp$ and $nn$ pairs, as predicted by the YPB laws. These correlations hold true for all chosen Skyrme parameter sets. 
We then extend our calculations across the nuclear chart, covering nuclei with mass number $40<A<240$, and investigate the generalized Levinger constants. The generalized Levinger constants are almost unchanged for both spin-zero $nn$ and $pp$ pairs, which implies a universal relationship between nuclear contacts and nuclear radii. This indicates that the likelihood of finding SRC pairs is proportional to the nucleon number density, and hence is a long-range, mean-field quantity. 
The globally determined generalized Levinger constants provide a robust tool for evaluating nuclear contacts based on nuclear radii across the chart of nuclides, which can be extracted from the elastic electron scattering experiments for both stable and unstable nuclei.

The high-momentum tails of SRC pairs show obvious impacts on the equation of state of neutron stars~\cite{Frankfurt2008,Li2018lpy}. In neutron stars, the abundance of SRC pairs is commonly estimated to be around 20\%, similar to what is observed in quasi-elastic electron scattering experiments off finite nuclei~\cite{Wang2021gse,Wang2023}. However, according to the analyses in this work, the formation of spin-zero $nn$ and $pp$ pairs is strongly influenced by the nucleon number density. Taking this effect into account, the abundance of SRC pairs in neutron stars may differ from that in finite nuclei, as the core density of neutron stars is several times higher than nuclear saturation density.

\section*{Acknowledgements}
This work is supported by the National Natural Science Foundation of China (Grants No.\ 12035011, No.\ 11975167,  No.\ 11947211, No.\ 11905103,  No.\  11881240623, No.\ 11961141003, and No.\ 12375122), and by the National Key R\&D Program of China (Contracts No.\ 2023YFA1606503 and No.\ 2018YFA0404403).

\appendix
\section{Supplemental Materials}
\setcounter{figure}{0} 

\begin{figure}[h] \centering
\includegraphics[width=0.8\columnwidth]{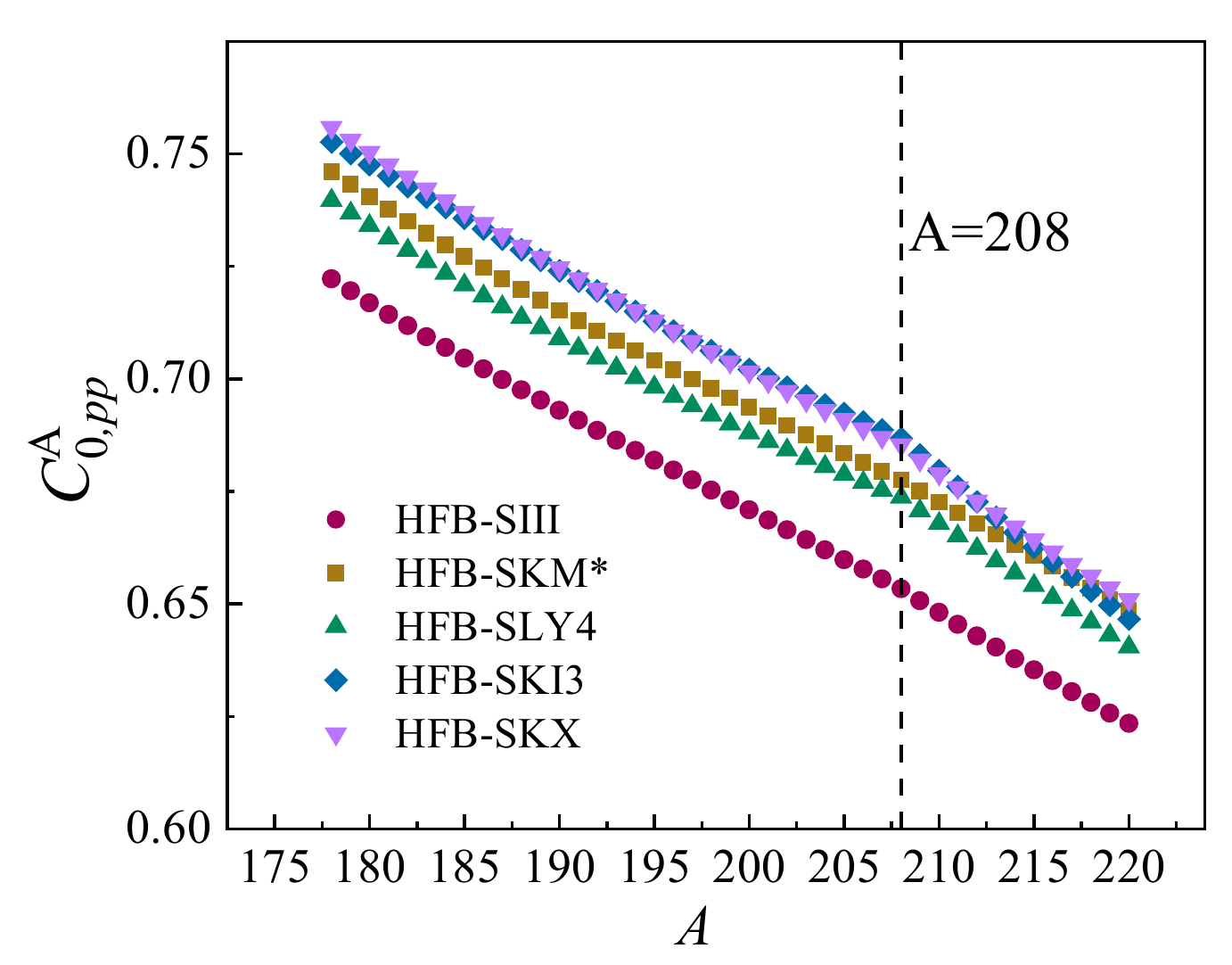}
\caption{The $pp$ contacts $C^A_{0,pp}$ for Pb isotopes versus mass number $A$. The corresponding point-proton density distributions are given by the Skyrme HFB method with SIII, SKM*, SLY4, SKI3, and SKX parameter sets. The vertical dashed line marks $A=208$.}
\label{Pb-A}
\end{figure}

\begin{figure*}[h] \centering
\subfigure{
\label{82:a}	\includegraphics[width=0.8\columnwidth]{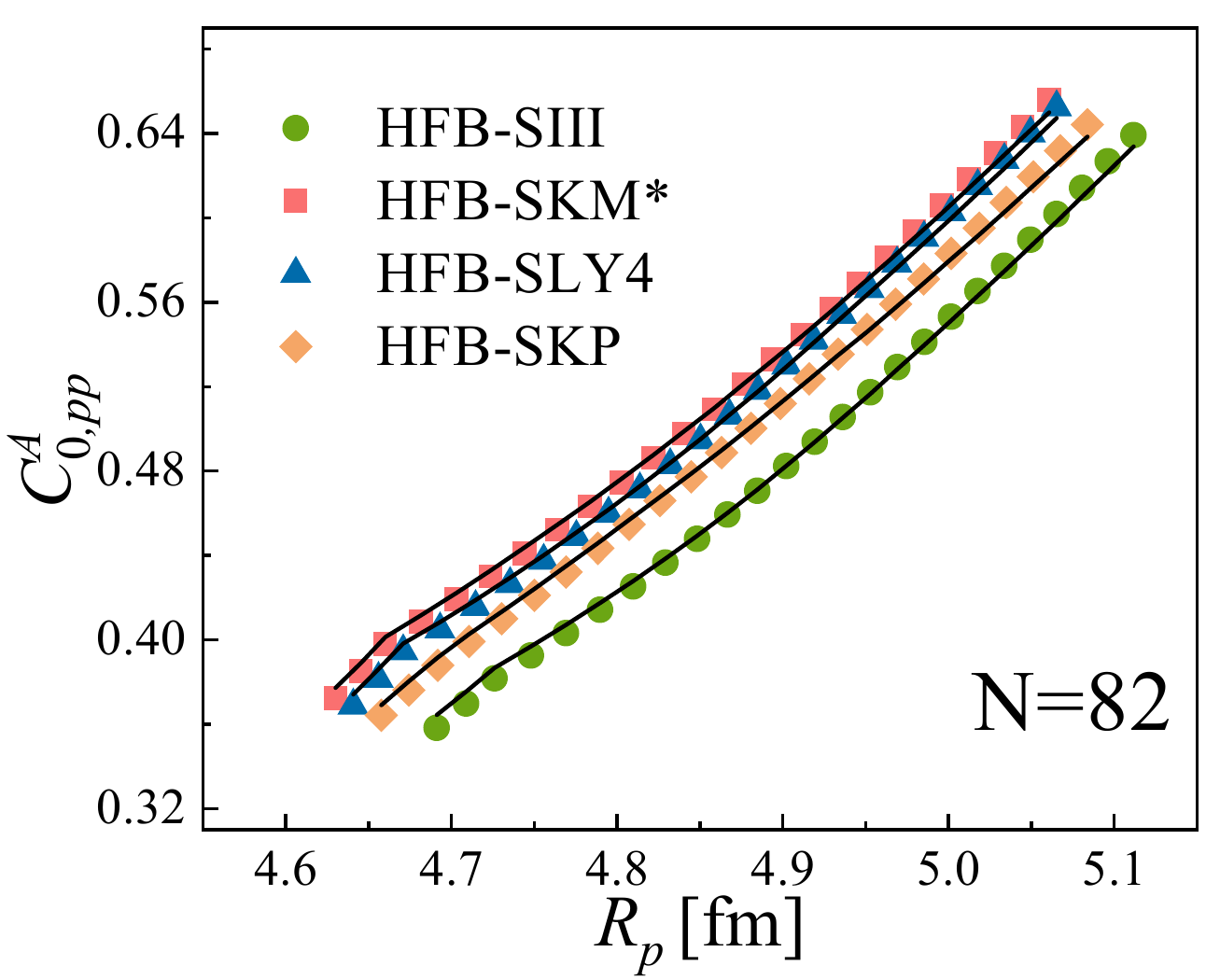}
		}
\subfigure {
\label{82:b}
\includegraphics[width=0.8\columnwidth]{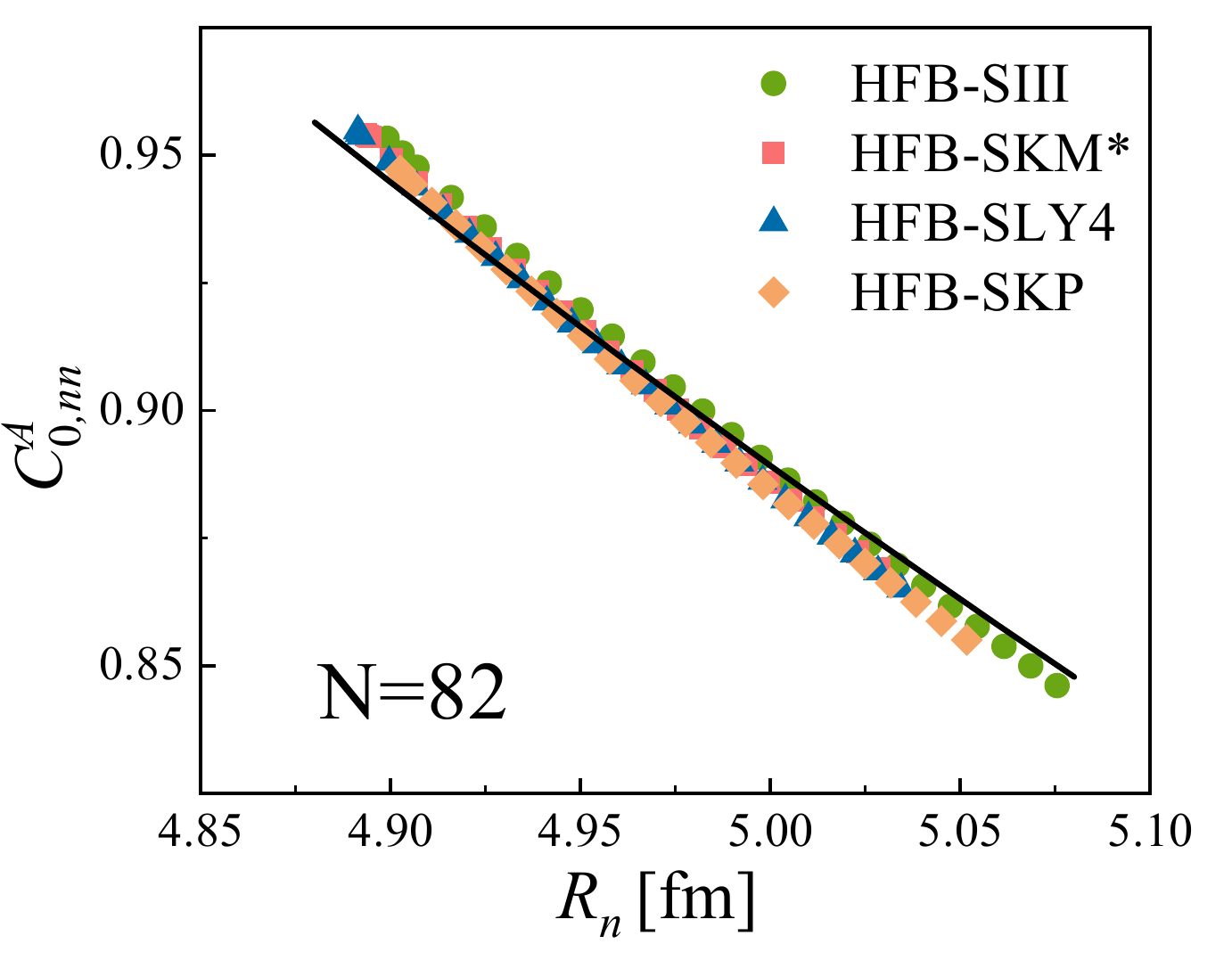}
		}
\caption{Nuclear contacts for $N=82$ isotones versus RMS radii of nucleon density distributions for (a) $pp$ pairs and (b) $nn$ pairs. The results are shown for nuclei with proton numbers ranging from $Z=48$ to $Z=72$. The fitted lines follow the expressions $C^A_{0,pp}=L^0_{pp}\frac{Z^2}{R_p^3}$ and $C^A_{0,nn}=L^0_{nn}\frac{82^2}{R_n^3}$. 
 The corresponding point-nucleon density distributions are given by the Skyrme HFB method with SIII, SKM*, SLY4, and SKP parameter sets. 
 The generalized Levinger constant is determined to be $ L^0_{nn} = 0.01653$ fm$^3$ for $nn$ pairs, while for $pp$ pairs, the values of $ L^0_{pp} $ range from 0.01618 to 0.01633 fm$^3$ across different data sets.}
\label{82}
\end{figure*}

\begin{figure*}[h] \centering
\includegraphics[width=1.7\columnwidth]{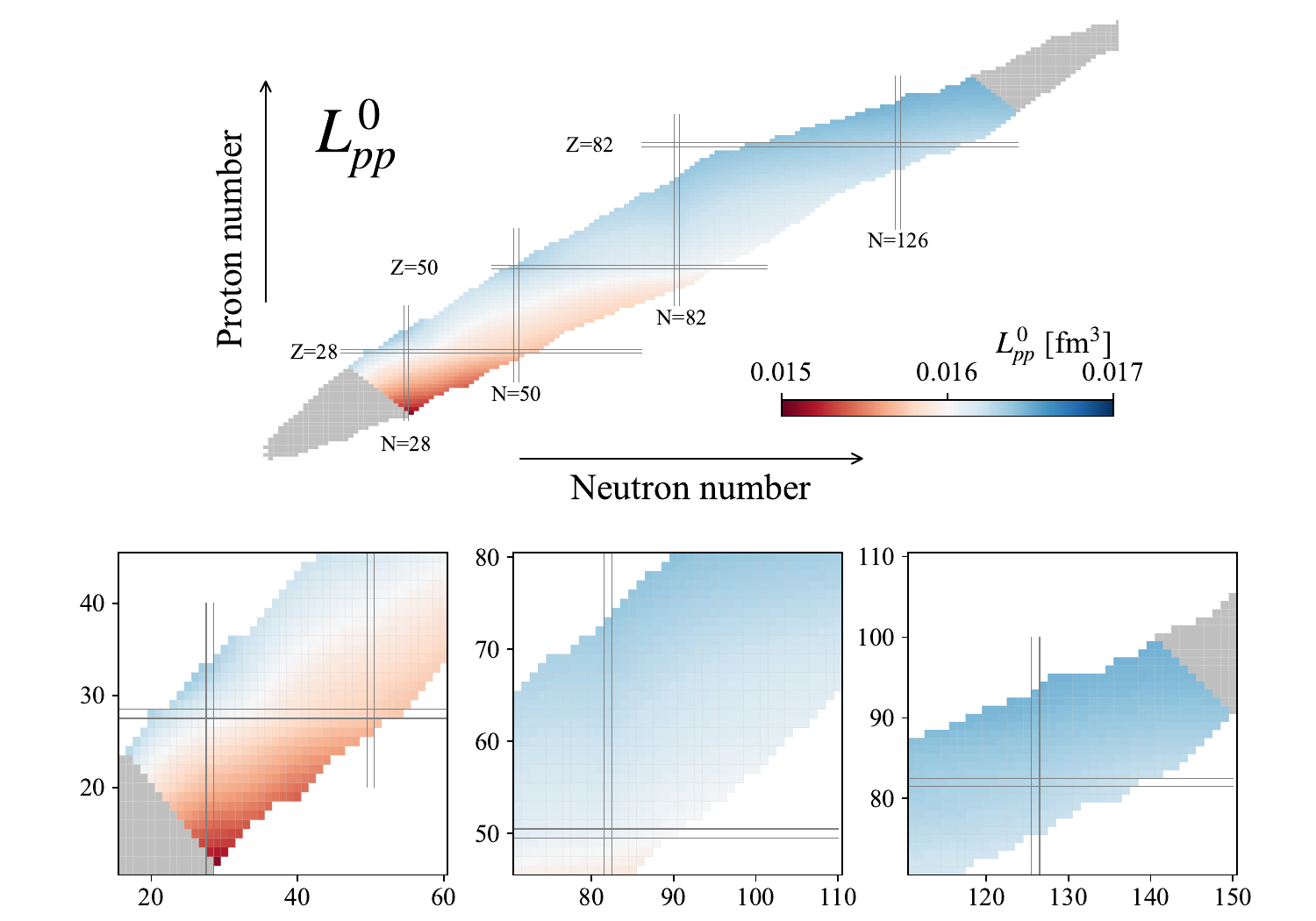}
\caption{Generalized Levinger constant for $pp$ pairs $L^0_{pp}$ across the chart of nuclides. The three images below display an enlarged view of the pattern. The corresponding $pp$ contacts and RMS radii of proton density distributions are calculated with the HFB point-proton density distributions using the SLY4 parameter set.}
\label{L_pp}
\end{figure*}

\begin{figure*}[h] \centering
\includegraphics[width=0.9\columnwidth]{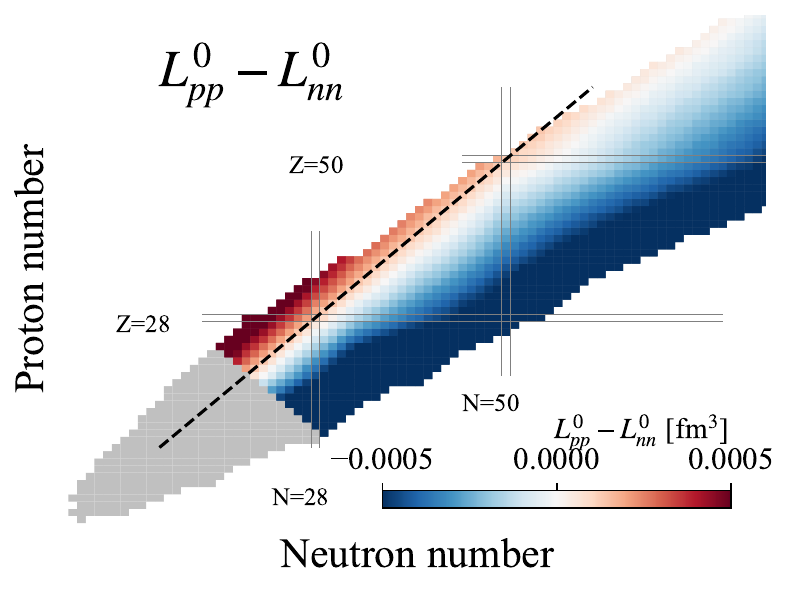}
\caption{Difference between $L^0_{pp}$ and $L^0_{nn}$ for nuclei $40<A<80$. }
\label{L_pp-L_nn}
\end{figure*}

\bibliographystyle{elsarticle-num-names}
\bibliography{references}

\begin{thebibliography}{48}
\expandafter\ifx\csname natexlab\endcsname\relax\def\natexlab#1{#1}\fi
\providecommand{\url}[1]{\texttt{#1}}
\providecommand{\href}[2]{#2}
\providecommand{\path}[1]{#1}
\providecommand{\DOIprefix}{doi:}
\providecommand{\ArXivprefix}{arXiv:}
\providecommand{\URLprefix}{URL: }
\providecommand{\Pubmedprefix}{pmid:}
\providecommand{\doi}[1]{\href{http://dx.doi.org/#1}{\path{#1}}}
\providecommand{\Pubmed}[1]{\href{pmid:#1}{\path{#1}}}
\providecommand{\bibinfo}[2]{#2}
\ifx\xfnm\relax \def\xfnm[#1]{\unskip,\space#1}\fi
\bibitem[{Ciofi~degli Atti(2015)}]{Ciofi2015}
\bibinfo{author}{C.~Ciofi~degli Atti},
\newblock \bibinfo{title}{In-medium short-range dynamics of nucleons: Recent
  theoretical and experimental advances},
\newblock \bibinfo{journal}{Phys. Rep.} \bibinfo{volume}{590}
  (\bibinfo{year}{2015}) \bibinfo{pages}{1--85}.
  \DOIprefix\doi{10.1016/j.physrep.2015.06.002}.
\bibitem[{Liang et~al.(2022)Liang, Ren, Bai, and Liu}]{Liang2022}
\bibinfo{author}{T.~Liang}, \bibinfo{author}{Z.~Ren}, \bibinfo{author}{D.~Bai},
  \bibinfo{author}{J.~Liu},
\newblock \bibinfo{title}{Nucleon momentum distributions from inclusive
  electron scattering with superscaling analysis},
\newblock \bibinfo{journal}{Phys. Rev. C} \bibinfo{volume}{106}
  (\bibinfo{year}{2022}) \bibinfo{pages}{054324}.
  \DOIprefix\doi{10.1103/PhysRevC.106.054324}.
\bibitem[{Niu et~al.(2022)Niu, Liu, Guo, Xu, Lyu, and Ren}]{Niu2022}
\bibinfo{author}{Q.~Niu}, \bibinfo{author}{J.~Liu}, \bibinfo{author}{Y.~Guo},
  \bibinfo{author}{C.~Xu}, \bibinfo{author}{M.~Lyu}, \bibinfo{author}{Z.~Ren},
\newblock \bibinfo{title}{Effects of nucleon-nucleon short-range correlations
  on inclusive electron scattering},
\newblock \bibinfo{journal}{Phys. Rev. C} \bibinfo{volume}{105}
  (\bibinfo{year}{2022}) \bibinfo{pages}{L051602}. \URLprefix
  \url{https://link.aps.org/doi/10.1103/PhysRevC.105.L051602}.
  \DOIprefix\doi{10.1103/PhysRevC.105.L051602}.
\bibitem[{Liang et~al.(2024)Liang, Bai, and Ren}]{Liang20241}
\bibinfo{author}{T.~Liang}, \bibinfo{author}{D.~Bai}, \bibinfo{author}{Z.~Ren},
\newblock \bibinfo{title}{Modern numerical differentiation technique for
  extracting nucleon momentum distributions},
\newblock \bibinfo{journal}{Phys. Rev. C} \bibinfo{volume}{109}
  (\bibinfo{year}{2024}) \bibinfo{pages}{054313}.
  \DOIprefix\doi{10.1103/PhysRevC.109.054313}.
\bibitem[{Hen et~al.(2015)Hen, Li, Guo, Weinstein, and Piasetzky}]{Hen2015}
\bibinfo{author}{O.~Hen}, \bibinfo{author}{B.-A. Li}, \bibinfo{author}{W.-J.
  Guo}, \bibinfo{author}{L.~B. Weinstein}, \bibinfo{author}{E.~Piasetzky},
\newblock \bibinfo{title}{Symmetry energy of nucleonic matter with tensor
  correlations},
\newblock \bibinfo{journal}{Phys. Rev. C} \bibinfo{volume}{91}
  (\bibinfo{year}{2015}) \bibinfo{pages}{025803}.
  \DOIprefix\doi{10.1103/PhysRevC.91.025803}.
\bibitem[{Cai and Li(2016)}]{cai2016}
\bibinfo{author}{B.-J. Cai}, \bibinfo{author}{B.-A. Li},
\newblock \bibinfo{title}{Symmetry energy of cold nucleonic matter within a
  relativistic mean field model encapsulating effects of high-momentum nucleons
  induced by short-range correlations},
\newblock \bibinfo{journal}{Phys. Rev. C} \bibinfo{volume}{93}
  (\bibinfo{year}{2016}) \bibinfo{pages}{014619}.
  \DOIprefix\doi{10.1103/PhysRevC.93.014619}.
\bibitem[{Lu et~al.(2022)Lu, Ren, and Bai}]{Lu2022}
\bibinfo{author}{H.~Lu}, \bibinfo{author}{Z.~Ren}, \bibinfo{author}{D.~Bai},
\newblock \bibinfo{title}{Neutron-neutron short-range correlations and their
  impacts on neutron stars},
\newblock \bibinfo{journal}{Nucl. Phys. A} \bibinfo{volume}{1021}
  (\bibinfo{year}{2022}) \bibinfo{pages}{122408}.
  \DOIprefix\doi{10.1016/j.nuclphysa.2022.122408}.
\bibitem[{Hong et~al.(2023)Hong, Ren, Wu, and Mu}]{Hong2023}
\bibinfo{author}{B.~Hong}, \bibinfo{author}{Z.~Ren}, \bibinfo{author}{C.~Wu},
  \bibinfo{author}{X.~Mu},
\newblock \bibinfo{title}{Impacts of symmetry energy slope on the oscillation
  frequencies of neutron stars with short-range correlation and admixed dark
  matter},
\newblock \bibinfo{journal}{Class. Quantum Grav.} \bibinfo{volume}{40}
  (\bibinfo{year}{2023}) \bibinfo{pages}{125007}.
  \DOIprefix\doi{10.1088/1361-6382/acd516}.
\bibitem[{Weinstein et~al.(2011)Weinstein, Piasetzky, Higinbotham, Gomez, Hen,
  and Shneor}]{Weinstein2011}
\bibinfo{author}{L.~B. Weinstein}, \bibinfo{author}{E.~Piasetzky},
  \bibinfo{author}{D.~W. Higinbotham}, \bibinfo{author}{J.~Gomez},
  \bibinfo{author}{O.~Hen}, \bibinfo{author}{R.~Shneor},
\newblock \bibinfo{title}{Short range correlations and the emc effect},
\newblock \bibinfo{journal}{Phys. Rev. Lett.} \bibinfo{volume}{106}
  (\bibinfo{year}{2011}) \bibinfo{pages}{052301}.
  \DOIprefix\doi{10.1103/PhysRevLett.106.052301}.
\bibitem[{Hen et~al.(2012)Hen, Piasetzky, and Weinstein}]{Hen2012}
\bibinfo{author}{O.~Hen}, \bibinfo{author}{E.~Piasetzky},
  \bibinfo{author}{L.~B. Weinstein},
\newblock \bibinfo{title}{New data strengthen the connection between short
  range correlations and the emc effect},
\newblock \bibinfo{journal}{Phys. Rev. C} \bibinfo{volume}{85}
  (\bibinfo{year}{2012}) \bibinfo{pages}{047301}.
  \DOIprefix\doi{10.1103/PhysRevC.85.047301}.
\bibitem[{Hen et~al.(2017)Hen, Miller, Piasetzky, and Weinstein}]{Hen2017}
\bibinfo{author}{O.~Hen}, \bibinfo{author}{G.~A. Miller},
  \bibinfo{author}{E.~Piasetzky}, \bibinfo{author}{L.~B. Weinstein},
\newblock \bibinfo{title}{Nucleon-nucleon correlations, short-lived
  excitations, and the quarks within},
\newblock \bibinfo{journal}{Rev. Mod. Phys.} \bibinfo{volume}{89}
  (\bibinfo{year}{2017}) \bibinfo{pages}{045002}.
  \DOIprefix\doi{10.1103/RevModPhys.89.045002}.
\bibitem[{Chen et~al.(2017)Chen, Detmold, Lynn, and Schwenk}]{chen2017}
\bibinfo{author}{J.-W. Chen}, \bibinfo{author}{W.~Detmold},
  \bibinfo{author}{J.~E. Lynn}, \bibinfo{author}{A.~Schwenk},
\newblock \bibinfo{title}{Short-range correlations and the emc effect in
  effective field theory},
\newblock \bibinfo{journal}{Phys. Rev. Lett.} \bibinfo{volume}{119}
  (\bibinfo{year}{2017}) \bibinfo{pages}{262502}.
  \DOIprefix\doi{10.1103/PhysRevLett.119.262502}.
\bibitem[{\ifmmode~\check{S}\else \v{S}\fi{}imkovic
  et~al.(2009)\ifmmode~\check{S}\else \v{S}\fi{}imkovic, Faessler, M\"uther,
  Rodin, and Stauf}]{Simkovic2009}
\bibinfo{author}{F.~\ifmmode~\check{S}\else \v{S}\fi{}imkovic},
  \bibinfo{author}{A.~Faessler}, \bibinfo{author}{H.~M\"uther},
  \bibinfo{author}{V.~Rodin}, \bibinfo{author}{M.~Stauf},
\newblock
  \bibinfo{title}{$0\ensuremath{\nu}\ensuremath{\beta}\ensuremath{\beta}$-decay
  nuclear matrix elements with self-consistent short-range correlations},
\newblock \bibinfo{journal}{Phys. Rev. C} \bibinfo{volume}{79}
  (\bibinfo{year}{2009}) \bibinfo{pages}{055501}. \URLprefix
  \url{https://link.aps.org/doi/10.1103/PhysRevC.79.055501}.
  \DOIprefix\doi{10.1103/PhysRevC.79.055501}.
\bibitem[{Cirigliano et~al.(2018)Cirigliano, Dekens, de~Vries, Graesser,
  Mereghetti, Pastore, and van Kolck}]{Cirigliano2018}
\bibinfo{author}{V.~Cirigliano}, \bibinfo{author}{W.~Dekens},
  \bibinfo{author}{J.~de~Vries}, \bibinfo{author}{M.~L. Graesser},
  \bibinfo{author}{E.~Mereghetti}, \bibinfo{author}{S.~Pastore},
  \bibinfo{author}{U.~van Kolck},
\newblock \bibinfo{title}{New leading contribution to neutrinoless
  double-$\ensuremath{\beta}$ decay},
\newblock \bibinfo{journal}{Phys. Rev. Lett.} \bibinfo{volume}{120}
  (\bibinfo{year}{2018}) \bibinfo{pages}{202001}. \URLprefix
  \url{https://link.aps.org/doi/10.1103/PhysRevLett.120.202001}.
  \DOIprefix\doi{10.1103/PhysRevLett.120.202001}.
\bibitem[{Jokiniemi et~al.(2021)Jokiniemi, Soriano, and
  Menéndez}]{Jokiniemi2021}
\bibinfo{author}{L.~Jokiniemi}, \bibinfo{author}{P.~Soriano},
  \bibinfo{author}{J.~Menéndez},
\newblock \bibinfo{title}{Impact of the leading-order short-range nuclear
  matrix element on the neutrinoless double-beta decay of medium-mass and heavy
  nuclei},
\newblock \bibinfo{journal}{Phys. Lett. B} \bibinfo{volume}{823}
  (\bibinfo{year}{2021}) \bibinfo{pages}{136720}. \URLprefix
  \url{https://www.sciencedirect.com/science/article/pii/S0370269321006602}.
  \DOIprefix\doi{https://doi.org/10.1016/j.physletb.2021.136720}.
\bibitem[{Weiss et~al.(2022)Weiss, Soriano, Lovato, Menendez, and
  Wiringa}]{Weiss2022}
\bibinfo{author}{R.~Weiss}, \bibinfo{author}{P.~Soriano},
  \bibinfo{author}{A.~Lovato}, \bibinfo{author}{J.~Menendez},
  \bibinfo{author}{R.~B. Wiringa},
\newblock \bibinfo{title}{Neutrinoless double-$\ensuremath{\beta}$ decay:
  Combining quantum monte carlo and the nuclear shell model with the
  generalized contact formalism},
\newblock \bibinfo{journal}{Phys. Rev. C} \bibinfo{volume}{106}
  (\bibinfo{year}{2022}) \bibinfo{pages}{065501}. \URLprefix
  \url{https://link.aps.org/doi/10.1103/PhysRevC.106.065501}.
  \DOIprefix\doi{10.1103/PhysRevC.106.065501}.
\bibitem[{Tan(2008{\natexlab{a}})}]{TAN20082952}
\bibinfo{author}{S.~Tan},
\newblock \bibinfo{title}{Energetics of a strongly correlated fermi gas},
\newblock \bibinfo{journal}{Ann. Phys. (NY)} \bibinfo{volume}{323}
  (\bibinfo{year}{2008}{\natexlab{a}}) \bibinfo{pages}{2952--2970}.
  \DOIprefix\doi{https://doi.org/10.1016/j.aop.2008.03.004}.
\bibitem[{Tan(2008{\natexlab{b}})}]{TAN20082971}
\bibinfo{author}{S.~Tan},
\newblock \bibinfo{title}{Large momentum part of a strongly correlated fermi
  gas},
\newblock \bibinfo{journal}{Ann. Phys. (NY)} \bibinfo{volume}{323}
  (\bibinfo{year}{2008}{\natexlab{b}}) \bibinfo{pages}{2971--2986}.
  \DOIprefix\doi{https://doi.org/10.1016/j.aop.2008.03.005}.
\bibitem[{Tan(2008{\natexlab{c}})}]{TAN20082987}
\bibinfo{author}{S.~Tan},
\newblock \bibinfo{title}{Generalized virial theorem and pressure relation for
  a strongly correlated fermi gas},
\newblock \bibinfo{journal}{Ann. Phys. (NY)} \bibinfo{volume}{323}
  (\bibinfo{year}{2008}{\natexlab{c}}) \bibinfo{pages}{2987--2990}.
  \DOIprefix\doi{https://doi.org/10.1016/j.aop.2008.03.003}.
\bibitem[{Hen et~al.(2015)Hen, Weinstein, Piasetzky, Miller, Sargsian, and
  Sagi}]{Hen20152}
\bibinfo{author}{O.~Hen}, \bibinfo{author}{L.~B. Weinstein},
  \bibinfo{author}{E.~Piasetzky}, \bibinfo{author}{G.~A. Miller},
  \bibinfo{author}{M.~M. Sargsian}, \bibinfo{author}{Y.~Sagi},
\newblock \bibinfo{title}{Correlated fermions in nuclei and ultracold atomic
  gases},
\newblock \bibinfo{journal}{Phys. Rev. C} \bibinfo{volume}{92}
  (\bibinfo{year}{2015}) \bibinfo{pages}{045205}.
  \DOIprefix\doi{10.1103/PhysRevC.92.045205}.
\bibitem[{Weiss et~al.(2015)Weiss, Bazak, and Barnea}]{Weiss2015}
\bibinfo{author}{R.~Weiss}, \bibinfo{author}{B.~Bazak},
  \bibinfo{author}{N.~Barnea},
\newblock \bibinfo{title}{Generalized nuclear contacts and momentum
  distributions},
\newblock \bibinfo{journal}{Phys. Rev. C} \bibinfo{volume}{92}
  (\bibinfo{year}{2015}) \bibinfo{pages}{054311}.
  \DOIprefix\doi{10.1103/PhysRevC.92.054311}.
\bibitem[{Cruz-Torres et~al.(2020)Cruz-Torres, Lonardoni, Weiss, Piarulli,
  Barnea, Higinbotham, Piasetzky, Schmidt, Weinstein, Wiringa, and
  Hen}]{Cruz-Torres2020}
\bibinfo{author}{R.~Cruz-Torres}, \bibinfo{author}{D.~Lonardoni},
  \bibinfo{author}{R.~Weiss}, \bibinfo{author}{M.~Piarulli},
  \bibinfo{author}{N.~Barnea}, \bibinfo{author}{D.~W. Higinbotham},
  \bibinfo{author}{E.~Piasetzky}, \bibinfo{author}{A.~Schmidt},
  \bibinfo{author}{L.~B. Weinstein}, \bibinfo{author}{R.~B. Wiringa},
  \bibinfo{author}{O.~Hen},
\newblock \bibinfo{title}{Many-body factorization and position–momentum
  equivalence of nuclear short-range correlations},
\newblock \bibinfo{journal}{Nature Phys.} \bibinfo{volume}{17}
  (\bibinfo{year}{2020}) \bibinfo{pages}{306--310}.
  \DOIprefix\doi{10.1038/s41567-020-01053-7}.
\bibitem[{Weiss et~al.(2015)Weiss, Bazak, and Barnea}]{Weiss20152}
\bibinfo{author}{R.~Weiss}, \bibinfo{author}{B.~Bazak},
  \bibinfo{author}{N.~Barnea},
\newblock \bibinfo{title}{Nuclear neutron-proton contact and the
  photoabsorption cross section},
\newblock \bibinfo{journal}{Phys. Rev. Lett.} \bibinfo{volume}{114}
  (\bibinfo{year}{2015}) \bibinfo{pages}{012501}. \URLprefix
  \url{https://www.ncbi.nlm.nih.gov/pubmed/25615461}.
  \DOIprefix\doi{10.1103/PhysRevLett.114.012501}.
\bibitem[{Weiss et~al.(2018)Weiss, Cruz-Torres, Barnea, Piasetzky, and
  Hen}]{Weiss2018}
\bibinfo{author}{R.~Weiss}, \bibinfo{author}{R.~Cruz-Torres},
  \bibinfo{author}{N.~Barnea}, \bibinfo{author}{E.~Piasetzky},
  \bibinfo{author}{O.~Hen},
\newblock \bibinfo{title}{The nuclear contacts and short range correlations in
  nuclei},
\newblock \bibinfo{journal}{Phys. Lett. B} \bibinfo{volume}{780}
  (\bibinfo{year}{2018}) \bibinfo{pages}{211--215}.
  \DOIprefix\doi{10.1016/j.physletb.2018.01.061}.
\bibitem[{Patsyuk et~al.(2021)}]{Patsyuk2021}
\bibinfo{author}{M.~Patsyuk}, et~al.,
\newblock \bibinfo{title}{Unperturbed inverse kinematics nucleon knockout
  measurements with a carbon beam},
\newblock \bibinfo{journal}{Nature Phys.} \bibinfo{volume}{17}
  (\bibinfo{year}{2021}) \bibinfo{pages}{693--699}.
  \DOIprefix\doi{10.1038/s41567-021-01193-4}.
\bibitem[{Liang et~al.(2024)Liang, Bai, and Ren}]{Liang2024}
\bibinfo{author}{T.~Liang}, \bibinfo{author}{D.~Bai}, \bibinfo{author}{Z.~Ren},
\newblock \bibinfo{title}{Nuclear contacts of unstable nuclei},
\newblock \bibinfo{journal}{Phys. Lett. B} \bibinfo{volume}{857}
  (\bibinfo{year}{2024}) \bibinfo{pages}{138965}. \URLprefix
  \url{https://www.sciencedirect.com/science/article/pii/S0370269324005239}.
  \DOIprefix\doi{https://doi.org/10.1016/j.physletb.2024.138965}.
\bibitem[{Yankovich et~al.(2024)Yankovich, Pazy, and Barnea}]{Yankovich2024}
\bibinfo{author}{R.~Yankovich}, \bibinfo{author}{E.~Pazy},
  \bibinfo{author}{N.~Barnea},
\newblock \bibinfo{title}{The relative abundance of correlated spin-zero
  nucleon pairs},
\newblock \bibinfo{journal}{arXiv e-prints}  (\bibinfo{year}{2024}). \URLprefix
  \url{https://ui.adsabs.harvard.edu/abs/2024arXiv240715068Y}.
  \DOIprefix\doi{10.48550/arXiv.2407.15068}.
\bibitem[{Centelles et~al.(2009)Centelles, Roca-Maza, Vinas, and
  Warda}]{Centelles2009}
\bibinfo{author}{M.~Centelles}, \bibinfo{author}{X.~Roca-Maza},
  \bibinfo{author}{X.~Vinas}, \bibinfo{author}{M.~Warda},
\newblock \bibinfo{title}{Nuclear symmetry energy probed by neutron skin
  thickness of nuclei},
\newblock \bibinfo{journal}{Phys. Rev. Lett.} \bibinfo{volume}{102}
  (\bibinfo{year}{2009}) \bibinfo{pages}{122502}. \URLprefix
  \url{https://www.ncbi.nlm.nih.gov/pubmed/19392269}.
  \DOIprefix\doi{10.1103/PhysRevLett.102.122502}.
\bibitem[{Adhikari et~al.(2021)}]{Adhikari2021}
\bibinfo{author}{D.~Adhikari}, et~al. (\bibinfo{collaboration}{PREX
  Collaboration}),
\newblock \bibinfo{title}{Accurate determination of the neutron skin thickness
  of $^{208}\mathrm{Pb}$ through parity-violation in electron scattering},
\newblock \bibinfo{journal}{Phys. Rev. Lett.} \bibinfo{volume}{126}
  (\bibinfo{year}{2021}) \bibinfo{pages}{172502}. \URLprefix
  \url{https://link.aps.org/doi/10.1103/PhysRevLett.126.172502}.
  \DOIprefix\doi{10.1103/PhysRevLett.126.172502}.
\bibitem[{Adhikari et~al.(2022)}]{Adhikari2022}
\bibinfo{author}{D.~Adhikari}, et~al. (\bibinfo{collaboration}{CREX
  Collaboration}),
\newblock \bibinfo{title}{Precision determination of the neutral weak form
  factor of $^{48}\mathrm{Ca}$},
\newblock \bibinfo{journal}{Phys. Rev. Lett.} \bibinfo{volume}{129}
  (\bibinfo{year}{2022}) \bibinfo{pages}{042501}. \URLprefix
  \url{https://link.aps.org/doi/10.1103/PhysRevLett.129.042501}.
  \DOIprefix\doi{10.1103/PhysRevLett.129.042501}.
\bibitem[{Reinhard et~al.(2021)Reinhard, Roca-Maza, and
  Nazarewicz}]{Reinhard2021}
\bibinfo{author}{P.~G. Reinhard}, \bibinfo{author}{X.~Roca-Maza},
  \bibinfo{author}{W.~Nazarewicz},
\newblock \bibinfo{title}{Information content of the parity-violating asymmetry
  in $^{208}\mathrm{Pb}$},
\newblock \bibinfo{journal}{Phys. Rev. Lett.} \bibinfo{volume}{127}
  (\bibinfo{year}{2021}) \bibinfo{pages}{232501}. \URLprefix
  \url{https://www.ncbi.nlm.nih.gov/pubmed/34936797}.
  \DOIprefix\doi{10.1103/PhysRevLett.127.232501}.
\bibitem[{Reinhard et~al.(2022)Reinhard, Roca-Maza, and
  Nazarewicz}]{Reinhard2022}
\bibinfo{author}{P.-G. Reinhard}, \bibinfo{author}{X.~Roca-Maza},
  \bibinfo{author}{W.~Nazarewicz},
\newblock \bibinfo{title}{Combined theoretical analysis of the parity-violating
  asymmetry for $^{48}\mathrm{Ca}$ and $^{208}\mathrm{Pb}$},
\newblock \bibinfo{journal}{Phys. Rev. Lett.} \bibinfo{volume}{129}
  (\bibinfo{year}{2022}) \bibinfo{pages}{232501}. \URLprefix
  \url{https://link.aps.org/doi/10.1103/PhysRevLett.129.232501}.
  \DOIprefix\doi{10.1103/PhysRevLett.129.232501}.
\bibitem[{Liang et~al.(2023)Liang, Wang, Liu, and Ren}]{Liang2023}
\bibinfo{author}{T.~Liang}, \bibinfo{author}{H.~Wang},
  \bibinfo{author}{J.~Liu}, \bibinfo{author}{Z.~Ren},
\newblock \bibinfo{title}{{Investigation of the neutron distribution
  deformation by parity-violating electron scattering}},
\newblock \bibinfo{journal}{Phys. Rev. C} \bibinfo{volume}{108}
  (\bibinfo{year}{2023}) \bibinfo{pages}{014312}.
  \DOIprefix\doi{10.1103/PhysRevC.108.014312}.
\bibitem[{Weiss et~al.(2019)Weiss, Schmidt, Miller, and Barnea}]{Weiss2019}
\bibinfo{author}{R.~Weiss}, \bibinfo{author}{A.~Schmidt},
  \bibinfo{author}{G.~A. Miller}, \bibinfo{author}{N.~Barnea},
\newblock \bibinfo{title}{Short-range correlations and the charge density},
\newblock \bibinfo{journal}{Phys. Lett. B} \bibinfo{volume}{790}
  (\bibinfo{year}{2019}) \bibinfo{pages}{484--489}.
  \DOIprefix\doi{10.1016/j.physletb.2019.01.053}.
\bibitem[{Hu et~al.(2022)}]{Hu2022}
\bibinfo{author}{B.~Hu}, et~al.,
\newblock \bibinfo{title}{{Ab initio predictions link the neutron skin of
  $^{208}$Pb to nuclear forces}},
\newblock \bibinfo{journal}{Nature Phys.} \bibinfo{volume}{18}
  (\bibinfo{year}{2022}) \bibinfo{pages}{1196--1200}.
  \DOIprefix\doi{10.1038/s41567-023-02324-9}.
  \href{http://arxiv.org/abs/2112.01125}{{\tt arXiv:2112.01125}}.
\bibitem[{Lonardoni et~al.(2017)Lonardoni, Lovato, Pieper, and
  Wiringa}]{Lonardoni2017}
\bibinfo{author}{D.~Lonardoni}, \bibinfo{author}{A.~Lovato},
  \bibinfo{author}{S.~C. Pieper}, \bibinfo{author}{R.~B. Wiringa},
\newblock \bibinfo{title}{{Variational calculation of the ground state of
  closed-shell nuclei up to A=40}},
\newblock \bibinfo{journal}{Phys. Rev. C} \bibinfo{volume}{96}
  (\bibinfo{year}{2017}) \bibinfo{pages}{024326}. \URLprefix
  \url{https://link.aps.org/doi/10.1103/PhysRevC.96.024326}.
  \DOIprefix\doi{10.1103/PhysRevC.96.024326}.
\bibitem[{Cruz-Torres et~al.(2018)Cruz-Torres, Schmidt, Miller, Weinstein,
  Barnea, Weiss, Piasetzky, and Hen}]{Cruz2018}
\bibinfo{author}{R.~Cruz-Torres}, \bibinfo{author}{A.~Schmidt},
  \bibinfo{author}{G.~A. Miller}, \bibinfo{author}{L.~B. Weinstein},
  \bibinfo{author}{N.~Barnea}, \bibinfo{author}{R.~Weiss},
  \bibinfo{author}{E.~Piasetzky}, \bibinfo{author}{O.~Hen},
\newblock \bibinfo{title}{Short range correlations and the isospin dependence
  of nuclear correlation functions},
\newblock \bibinfo{journal}{Phys. Lett. B} \bibinfo{volume}{785}
  (\bibinfo{year}{2018}) \bibinfo{pages}{304--308}.
  \DOIprefix\doi{10.1016/j.physletb.2018.07.069}.
\bibitem[{Negele and Vautherin(1972)}]{Negele1972}
\bibinfo{author}{J.~W. Negele}, \bibinfo{author}{D.~Vautherin},
\newblock \bibinfo{title}{Density-matrix expansion for an effective nuclear
  hamiltonian},
\newblock \bibinfo{journal}{Phys. Rev. C} \bibinfo{volume}{5}
  (\bibinfo{year}{1972}) \bibinfo{pages}{1472--1493}.
  \DOIprefix\doi{10.1103/PhysRevC.5.1472}.
\bibitem[{Stoitsov et~al.(2005)Stoitsov, Dobaczewski, Nazarewicz, and
  Ring}]{Stoitsov2005}
\bibinfo{author}{M.~V. Stoitsov}, \bibinfo{author}{J.~Dobaczewski},
  \bibinfo{author}{W.~Nazarewicz}, \bibinfo{author}{P.~Ring},
\newblock \bibinfo{title}{Axially deformed solution of the
  $\mathrm{S}$kyrme–$\mathrm{H}$artree–$\mathrm{F}$ock–$\mathrm{B}$ogolyubov
  equations using the transformed harmonic oscillator basis. the program hfbtho
  (v1.66p)},
\newblock \bibinfo{journal}{Comput. Phys. Commun.} \bibinfo{volume}{167}
  (\bibinfo{year}{2005}) \bibinfo{pages}{43--63}.
  \DOIprefix\doi{10.1016/j.cpc.2005.01.001}.
\bibitem[{Sarriguren et~al.(2007)Sarriguren, Gaidarov, Guerra, and
  Antonov}]{Sarriguren2007}
\bibinfo{author}{P.~Sarriguren}, \bibinfo{author}{M.~K. Gaidarov},
  \bibinfo{author}{E.~M.~d. Guerra}, \bibinfo{author}{A.~N. Antonov},
\newblock \bibinfo{title}{Nuclear skin emergence in skyrme deformed
  $\mathrm{H}$artree-$\mathrm{F}$ock calculations},
\newblock \bibinfo{journal}{Phys. Rev. C} \bibinfo{volume}{76}
  (\bibinfo{year}{2007}) \bibinfo{pages}{044322}.
  \DOIprefix\doi{10.1103/PhysRevC.76.044322}.
\bibitem[{Wang et~al.(2020)Wang, Liu, Liang, Ren, Xu, and Wang}]{Wang2020}
\bibinfo{author}{L.~Wang}, \bibinfo{author}{J.~Liu},
  \bibinfo{author}{T.~Liang}, \bibinfo{author}{Z.~Ren},
  \bibinfo{author}{C.~Xu}, \bibinfo{author}{S.~Wang},
\newblock \bibinfo{title}{{Charge form factors of exotic nuclei in deformed
  Hartree\textendash{}Fock\textendash{}Bogolyubov calculations}},
\newblock \bibinfo{journal}{J. Phys. G} \bibinfo{volume}{47}
  (\bibinfo{year}{2020}) \bibinfo{pages}{025105}.
  \DOIprefix\doi{10.1088/1361-6471/ab5325}.
\bibitem[{Marevic et~al.(2022)Marevic, Schunck, Ney, P\'erez, Verriere, and
  O'Neal}]{Marevic2021}
\bibinfo{author}{P.~Marevic}, \bibinfo{author}{N.~Schunck},
  \bibinfo{author}{E.~M. Ney}, \bibinfo{author}{R.~N. P\'erez},
  \bibinfo{author}{M.~Verriere}, \bibinfo{author}{J.~O'Neal},
\newblock \bibinfo{title}{{Axially-deformed solution of the
  Skyrme-Hartree-Fock-Bogoliubov equations using the transformed harmonic
  oscillator basis (IV) hfbtho (v4.0): A new version of the program}},
\newblock \bibinfo{journal}{Comput. Phys. Commun.} \bibinfo{volume}{276}
  (\bibinfo{year}{2022}) \bibinfo{pages}{108367}.
  \DOIprefix\doi{10.1016/j.cpc.2022.108367}.
\bibitem[{Hen et~al.(2014)}]{Hen2014}
\bibinfo{author}{O.~Hen}, et~al.,
\newblock \bibinfo{title}{Momentum sharing in imbalanced fermi systems},
\newblock \bibinfo{journal}{Science} \bibinfo{volume}{346}
  (\bibinfo{year}{2014}) \bibinfo{pages}{614--617}. \URLprefix
  \url{https://www.science.org/doi/abs/10.1126/science.1256785}.
  \DOIprefix\doi{doi:10.1126/science.1256785}.
\bibitem[{Korover et~al.(2021)}]{Korover2021}
\bibinfo{author}{I.~Korover}, et~al. (\bibinfo{collaboration}{CLAS}),
\newblock \bibinfo{title}{{12C(e,e'pN) measurements of short range correlations
  in the tensor-to-scalar interaction transition region}},
\newblock \bibinfo{journal}{Phys. Lett. B} \bibinfo{volume}{820}
  (\bibinfo{year}{2021}) \bibinfo{pages}{136523}.
  \DOIprefix\doi{10.1016/j.physletb.2021.136523}.
  \href{http://arxiv.org/abs/2004.07304}{{\tt arXiv:2004.07304}}.
\bibitem[{Frankfurt et~al.(2008)Frankfurt, Sargsian, and
  Strikman}]{Frankfurt2008}
\bibinfo{author}{L.~Frankfurt}, \bibinfo{author}{M.~Sargsian},
  \bibinfo{author}{M.~Strikman},
\newblock \bibinfo{title}{{Recent observation of short range nucleon
  correlations in nuclei and their implications for the structure of nuclei and
  neutron stars}},
\newblock \bibinfo{journal}{Int. J. Mod. Phys. A} \bibinfo{volume}{23}
  (\bibinfo{year}{2008}) \bibinfo{pages}{2991--3055}.
  \DOIprefix\doi{10.1142/S0217751X08041207}.
  \href{http://arxiv.org/abs/0806.4412}{{\tt arXiv:0806.4412}}.
\bibitem[{Li et~al.(2018)Li, Cai, Chen, and Xu}]{Li2018lpy}
\bibinfo{author}{B.-A. Li}, \bibinfo{author}{B.-J. Cai}, \bibinfo{author}{L.-W.
  Chen}, \bibinfo{author}{J.~Xu},
\newblock \bibinfo{title}{{Nucleon Effective Masses in Neutron-Rich Matter}},
\newblock \bibinfo{journal}{Prog. Part. Nucl. Phys.} \bibinfo{volume}{99}
  (\bibinfo{year}{2018}) \bibinfo{pages}{29--119}.
  \DOIprefix\doi{10.1016/j.ppnp.2018.01.001}.
  \href{http://arxiv.org/abs/1801.01213}{{\tt arXiv:1801.01213}}.
\bibitem[{Wang et~al.(2021)Wang, Niu, Zhang, Lyu, Liu, Xu, and
  Ren}]{Wang2021gse}
\bibinfo{author}{X.~Wang}, \bibinfo{author}{Q.~Niu},
  \bibinfo{author}{J.~Zhang}, \bibinfo{author}{M.~Lyu},
  \bibinfo{author}{J.~Liu}, \bibinfo{author}{C.~Xu}, \bibinfo{author}{Z.~Ren},
\newblock \bibinfo{title}{{Nucleon momentum distribution of $^{56}$Fe from the
  axially deformed relativistic mean-field model with nucleon-nucleon
  correlations}},
\newblock \bibinfo{journal}{Sci. China Phys. Mech. Astron.}
  \bibinfo{volume}{64} (\bibinfo{year}{2021}) \bibinfo{pages}{292011}.
  \DOIprefix\doi{10.1007/s11433-021-1729-5}.
  \href{http://arxiv.org/abs/2108.12087}{{\tt arXiv:2108.12087}}.
\bibitem[{Wang et~al.(2023)Wang, Niu, Zhang, Liu, and Ren}]{Wang2023}
\bibinfo{author}{L.~Wang}, \bibinfo{author}{Q.~Niu},
  \bibinfo{author}{J.~Zhang}, \bibinfo{author}{J.~Liu},
  \bibinfo{author}{Z.~Ren},
\newblock \bibinfo{title}{{New extended method for \ensuremath{\psi}' scaling
  function of inclusive electron scattering}},
\newblock \bibinfo{journal}{Sci. China Phys. Mech. Astron.}
  \bibinfo{volume}{66} (\bibinfo{year}{2023}) \bibinfo{pages}{102011}.
  \DOIprefix\doi{10.1007/s11433-023-2135-x}.

\end{thebibliography}
\end{document}